# Influences on Drivers' Understandings of Systems by Presenting Image Recognition Results

Bo Yang, Koichiro Inoue, Satoshi Kitazaki, and Kimihiko Nakano, *Member, IEEE*

*Abstract*— It is essential to help drivers have appropriate understandings of level 2 automated driving systems for keeping driving safety. A human machine interface (HMI) was proposed to present real time results of image recognition by the automated driving systems to drivers. It was expected that drivers could better understand the capabilities of the systems by observing the proposed HMI. Driving simulator experiments with 18 participants were preformed to evaluate the effectiveness of the proposed system. Experimental results indicated that the proposed HMI could effectively inform drivers of potential risks continuously and help drivers better understand the level 2 automated driving systems.

## I. INTRODUCTION

Level-2 automated driving systems have already been put into practical use in some limited conditions. The systems can automatedly perform several driving tasks to assists drivers, including adaptive cruise control (ACC), which automates acceleration and brake control to follow a leading car at a certain distance, and lane keeping assist system (LKAS), which automatically control steering to ensure the vehicle not to leave the lane. In some specific conditions, almost all driving tasks are performed by level 2 automated driving systems, however, the final responsibility for these driving tasks belongs to drivers. Drivers are always requested to pay attention to surrounding traffic situations, to prepare to take over the driving tasks if an emergency event occurs and exceeds the capability of the system.

Nowadays, level 2 automated driving systems are mainly available in simple traffic environments including highways, where few changes occur compared with urban streets. To make the automated driving available in more complex environment, it is important to help drivers have appropriate understandings of the systems. Drivers' understandings of systems may be concluded as "a rich and elaborate structure, reflecting the user's understanding of what the system contains, how it works, and why it works that way" or defined more precisely as "a representation of the typical causal interconnections involving actions and environmental events that influence the functioning of the system" [1], [2]. Drivers may have an inappropriate trust level on the systems if they do not appropriately understand the systems. Previous study doubted that highly automated systems may induce drivers to trust on the systems excessively [3]. Informing drivers of specifications, features and limitations of the automated driving systems is considered important to avoid over trust. If drivers understand the automated driving systems in detail and correctly, they may have safer operations in complicated driving environments which request intervention frequently.

However, it was reported that drivers could not have correct understandings about ADAS without any necessary supports. A previous study announced that about 70% ACC owners did not understand warnings from the systems about the limitations of the systems and many users trusted the systems excessively in some scenarios [4]. Another research also reported that about 40% users of vehicles did not read the owner's manual, while the rest would read only half of it [5]. Meanwhile, it was found that collisions occurred in some experiments due to driver's insufficient knowledge and inappropriate trust level of the systems [6]. Furthermore, a study focused on the evolution of the knowledge about ACC showed that experiencing some risky scenarios after the correct explanations about the systems in advance would help drivers obtain a good understanding of the system. However, another study doubted that the effectiveness was not enough and unsustainable for keeping driving safety even when some risks had been explained in advance [7].

Previous studies found that it might be effective to present the limitations, uncertainty, or the status of automated systems to drivers for safer driving operations [8–11]. It was observed that drivers relied appropriately on ACC and braked faster and more consistently when the time to collision, and time head way with the preceding vehicle were calculated and displayed for drivers, resulting in safe following distances and avoidance of collisions [8]. Meanwhile, drivers performed better in take-over situations when the uncertainty of the automated systems was calculated based on the forward visibility related to the amount of snowing and was displayed in real time for drivers [9]. However, these previous studies were normally focused on limited traffic conditions and were difficult to be applied in the environment with various risk factors.

This study proposes a human machine interface which presents recognition status of the system directly and continuously to drivers. Different types of risks could be presented, and drivers could be informed of various potential risks in complicated driving environment. It was expected that drivers' understandings of the characteristics of the systems could be promoted with the application of the proposed HMI, by continuously presenting the recognition status of the systems and the potential risks, resulting in safe operations.

This research was a part of a national project entitled "HMI and user education for Highly Automated Driving" supported by Cabinet Office, Government of Japan, SIP-adus, and funded by NEDO. The project was also conducted based on Germany-Japan Research Cooperation on Human Factors in Connected and Automated Driving entitled "CADJapanGermany: HF".

Bo Yang and Kimihiko Nakano are with the Institute of Industrial Science, the University of Tokyo, Tokyo, 153-8505, Japan (e-mail: b-yang@iis.u-tokyo.ac.jp; and knakano@iis.u-tokyo.ac.jp). Koichiro Inoue is with Graduate School of Engineering, the University of Tokyo, Tokyo, 153-8505, Japan (e-mail: k8o2u9@iis.u-tokyo.ac.jp).
Satoshi Kitazaki is with the Human-Centered Mobility Research Center, National Institute of Advanced Industrial Science and Technology, Ibaraki, 305-8566, Japan (e-mail: satoshi-kitazaki@aist.go.jp).

This paper will firstly introduce the proposed human machine interface and the detailed information of the experiment, including the experimental scenarios and conditions. Afterwards, the experimental results will be provided. Finally, the paper will be concluded with the implications of the proposed method.

## II. HUMAN MACHINE INTERFACE

Real time results of object detections for the forward traffic situations would be provided with the proposed HMI, as shown in Fig. 1. The results were displayed as bounding boxes surrounding each object which the system recognized. A head-up display fixed on the dashboard of the driving simulator was applied to present the results to drivers, as shown in Fig. 2. Drivers could interpret which objects could be recognized or overlooked by the system in real time and could understand the characteristics of the system by continuously observing the HMI.

Tensorflow Object Detection API and a pre-trained detection model were applied in the recognition system. The detection model consisted of Faster RCNN and ResNet and was trained with Microsoft COCO dataset. Meanwhile, to verify whether drivers could understand the recognition capabilities of the system or not, the recognition system was trained to be able to recognize cars, buses, and trucks, while could not recognize several objects, including motorcycles and pylons.

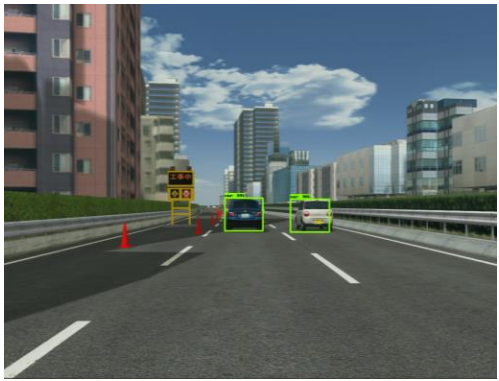

Fig. 1. Recognition results displayed to drivers, in which vehicles were recognized and pylons were not recognized.

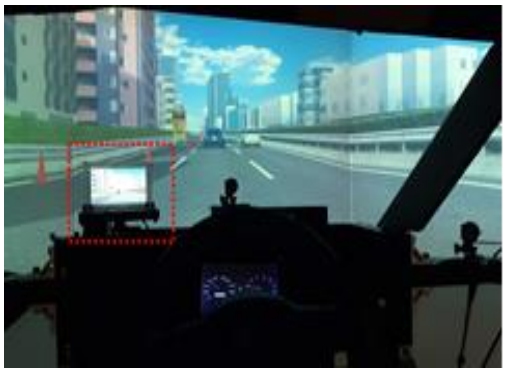

Fig. 2. Head-up display applied for the presentation of the results.

## III. EXPERIMENTS

### A. Participants

Eighteen participants, aged from 20 to 25 years old, attended the experiments. All participants had a valid driving license, and the experiments were conducted with the approval of the University of Tokyo.

### B. Apparatuses

A driving simulator equipped with three pairs of projectors and screens was applied, as presented in Fig. 3. Each screen provides the frontal forward, right forward, and left forward view, respectively. The frontal forward view generated by the host computer of the simulator was inputted into a notebook computer for performing object recognition. The resolution of inputted images was set as 1280*1024 pixels, and the system processed inputted images at 15 FPS and projected the results of object detection on the head-up display.

Meanwhile, level 2 automated driving function was simulated with the driving simulator. The simulated automated driving consisted of ACC and LKAS functions. The ACC function was achieved with the PID controller to converge distance between the ego and the leading vehicles into 20 meters and relative velocity into zero with a constant speed of 60km/h. The LKAS function worked as a simple controller which input a steering control signal corresponding to an angle between the direction of the ego vehicle and the direction from the ego vehicle to the leading vehicle. All necessary calculations for performing automated driving were executed by the host computer of the simulator.

### C. Experimental Scenario

The road environment used in the experiment is shown in Fig. 4, which reproduced the environment of Japanese major national roads. There are three lanes on each side in the first three quarters of whole scenario and the ego vehicle would drive on the second lane. Afterwards, the lanes were reduced to two lanes on each side in the rest of the scenario, and the lane where the ego vehicle was on would become the first lane. There were 28 intersections with non-priority roads while the ego vehicle was on the priority road for the whole scenario. During the experiment, lane change was not allowed unless emergency events occurred.

During the experiment, the automated driving functions were initially activated, and the participants did not need to perform driving tasks, however, they were required to operate manually to avoid collisions if risky situations occurred. Three types of potential risk scenes were prepared, including potential risks (a), (b) and (c), for the first three quarters of the scenario when there were three lanes, and each potential risk scene would appear several times. The potential risks would not result in collisions as they would not occur on the lane of the ego vehicle. For potential risk (a), another car would suddenly enter the first lane of the major road from a non-priority road after the leading car passed, which simulated a risky scene where ACC could not operate the brake in time, as shown in Fig. 5. The entering car was not recognized before it started to enter, however, no collision would occur as the ego vehicle was on the second lane. For the potential risk (b), pylons were applied to restrict the first lane, which were out of the recognition of the system, and no collision would occur, as presented in Fig. 6. For the potential risk (c), a motorcycle out of the recognition of the system would appear on the next lane of the ego vehicle, as provided in Fig. 7.

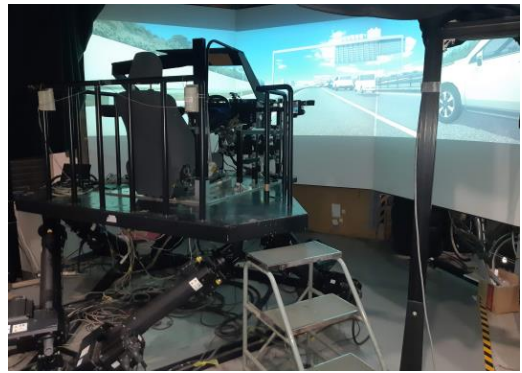

Fig. 3. Driving simulator.

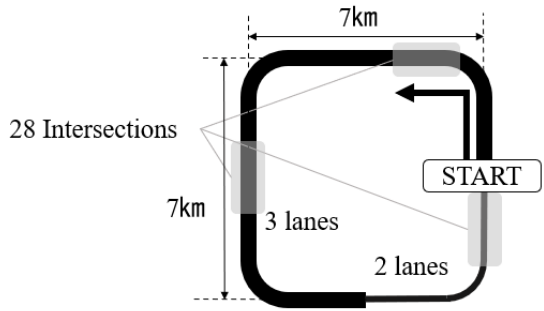

Fig. 4. Road environment.

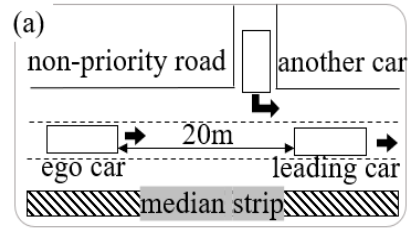

Fig. 5. Potential risk (a), in which another car would suddenly enter the first lane of the major road from a non-priority road after the leading car had passed.

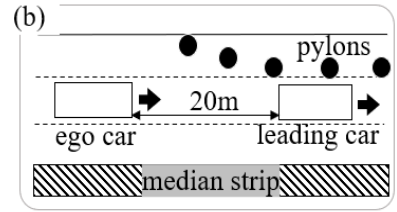

Fig. 6. Potential risk (b), in which pylons were applied to restrict the first lane, which were out of the recognition of the system.

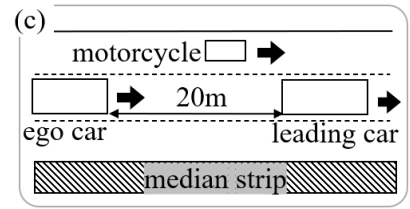

Fig. 7. Potential risk (c), in which a motorcycle would appear on the next lane of the ego vehicle, which was out of the recognition of the system.

Meanwhile, two types of apparent risk scenes, including apparent risks (A) and (B), were prepared for the last part of the scenario when there were two lanes on each side. For the apparent risk (A), another car would suddenly enter the lane of the ego vehicle from a non-priority road after the leading car passed, and the ego vehicle would collide with the entering car unless drivers braked manually, as shown in Fig. 8. For the apparent risk (B), pylons would suddenly appear on the lane of the ego vehicle after the leading vehicle changed the lane, and drivers needed to brake, or change lane manually to avoid collisions, as presented in Fig. 9.

Drivers would experience one apparent risk scene in one experimental condition. Therefore, two scenarios, including scenarios (i) and (ii), were prepared. The scenarios (i) consisted of potential risks (a), (b) and (c) and apparent risk (A). The scenarios (ii) consisted of potential risks (a), (b) and (c) and apparent risk (B). For both scenarios (i) and (ii), the number of the potential risks was the same, while the places where the apparent risk was shown varied in each scenario to avoid learning effects.

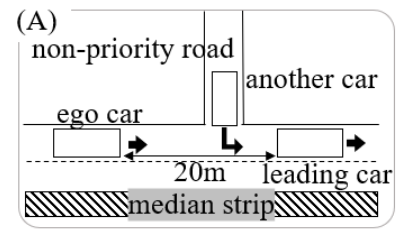

Fig. 8. Apparent risk (A), in which another car would suddenly enter the lane of the ego vehicle from a non-priority road after the leading car had passed.

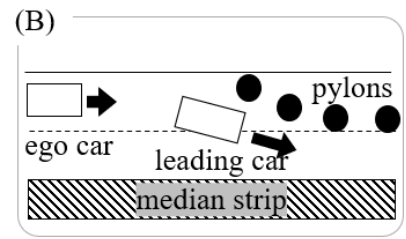

Fig. 9. Apparent risk (B), in which pylons would suddenly appear on the lane of the ego vehicle after the leading vehicle changed the lane.

*D. Experimental Conditions*

Eighteen participants were divided randomly into two groups, group (1) and group (2). For the participants in the group (1), they would experience the scenarios with the proposed HMI, while the participants in the group (2) would not use the HMI. The procedure of the experiment is shown in Fig. 10. Three questionnaires would be applied during the experiments, among which the questionnaire *A* would be used to obtain the basic information, including the driving experience, of the participants, and the questionnaire *B* would be applied to evaluate drivers' knowledge and understandings of the systems, as shown in Table I.

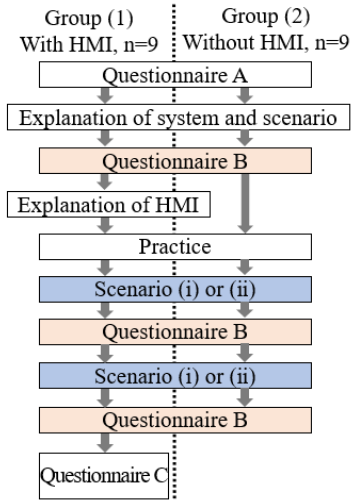

Fig. 10. Procedure of the experiments.

TABLE I. QUESTIONNAIRE *B*

| No. | Questions |
|---|---|
| Q1 | The system can recognize all vehicles |
| Q2 | The system can recognize motorcycle |
| Q3 | The system can recognize objects other than vehicles |
| Q4 | The system can recognize vehicles from non-priority road |
| Q5 | The system can recognize pedestrians |
| Q6 | The system can recognize bicycles |
| Q7 | The system can work safely without driver intervention |
| Q8 | The system can work safely for vehicles suddenly entering from the non-priority road side |
| Q9 | The system can work safely for pylons |
| Q10 | The system can work safely for pedestrians |
| Q11 | The system can work safely for bicycles |
| Q12 | The system can work safely for motorcycles |
| Q13 | There is no problem to look away from the surrounding traffic environment while the system is operating |
| Q14 | There is no problem to be more relaxed than manual operation while the system is operating |
| Q15 | There is no problem to use the smartphone while the system is operating |
| Q16 | There is no problem to get sleepy while the system is operating |
| Q17 | The system is reliable |
| Q18 | I would like to use the system on a daily basis |

According to the questionnaire *A*, there was no significant difference between the two groups in terms of driving experience. The mean driving experience was 37.2 months for the group (1) and 29.7 months for the group (2). As for the questionnaire *B*, it consisted of three parts, where Q1-Q6 were focused on the recognition of the system, Q7-Q12 on the reaction of the system, and Q13-Q18 on the level of trust for the system. Finally, questionnaire *C* would be used for the group (1) to evaluate the easiness of understanding the HMI. The questionnaires were conducted in five scales, where one for "strongly disagree" and five for "strongly agree".

As shown in Fig. 10, the information of the system and the scenarios were briefly introduced to the participants, after completing the questionnaire *A*. They were required to try their best to drive safely to the end of the scenario and prepare to take over if needed. However, the detailed information of the potential and apparent risks was not informed. Then the questionnaire *B* was performed to investigate drivers' initial knowledge and understandings of the system after the explanation. Afterwards, the explanations of HMI and the practice driving of the simulator were offered for the participants of group (1). As for the participants of the group (2), they only attended the driving practice. Then the scenario (i) and (ii) were provided for every participant, considering counterbalancing, and the evaluations with the questionnaire *B* were performed after each scenario.

IV. RESULTS

*A. Time to intervene*

The time to intervene index was defined as the time lag between the occurrence of apparent risk, and the timing when drivers intervened. The results were shown in Fig. 11. Shapiro-Wilk tests were firstly performed to check whether the data satisfy the assumption of normality. A nonparametric test, Wilcoxon rank sum test, was then performed, as the normality assumption was rejected by the Shapiro-Wilk test. It was found that no significant difference existed between the two groups for both the apparent risks (A) ($p = 0.56$) and (B) ($p = 0.64$). According to the feedbacks from the participants after the experiments, it was considered that some participants were observing the head-up display when apparent risks occurred, which might result in a delayed reaction even they quickly realized the risks and intended to intervene as soon as possible. Further investigations are needed to improve drivers' intervention performances.

*B. Subjective evaluations*

The results of subjective evaluations with the questionnaire *B* for groups (1) and (2) were presented with Figs. 12-14, where Fig. 12 offers the evaluations on the recognition of the system from Q1 to Q6, Fig. 13 for the evaluations on the reaction of the system from Q7 to Q12, and Fig. 14 for the level of trust for the system from Q13 to Q18. The words "first", "second" and "third" in the figures indicate the order of the questionnaire *B* as shown in Fig. 10. The leftmost of the figures are one point, indicating strongly disagree, and the rightmost are five points, indicating strongly agree. The *p* values were calculated with Wilcoxon rank sum test and provided in the Table II.

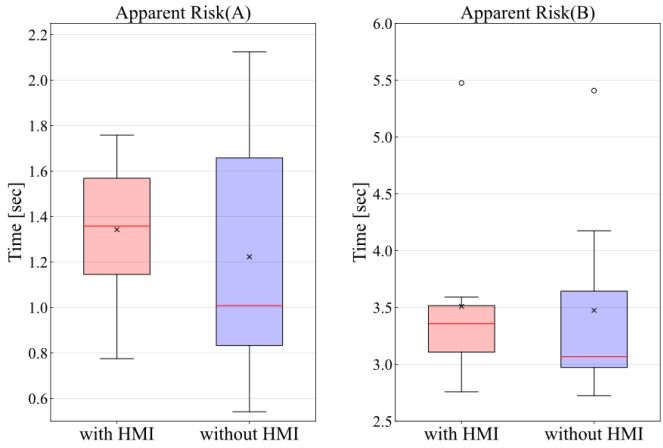

Fig. 11. Time to intervene. Mild outlier indicated by a circle (calculated as 1.5–3x the interquartile range).

TABLE II. RESULTS OF *P* VALUES

|     | first | second | third |
| --- | --- | --- | --- |
| Q1  | 0.83 | 0.58 | 0.41 |
| Q2  | 0.82 | **0.0029** | **0.00041** |
| Q3  | 0.20 | 0.064 | **0.045** |
| Q4  | 0.61 | 0.41 | **0.026** |
| Q5  | 0.36 | 0.79 | 0.73 |
| Q6  | 0.22 | **0.016** | **0.015** |
| Q7  | 0.068 | 0.062 | 1.0 |
| Q8  | 1.0 | 0.90 | 0.46 |
| Q9  | 0.36 | 0.43 | 0.23 |
| Q10 | 0.19 | 0.36 | 0.055 |
| Q11 | 0.30 | 0.10 | **0.0073** |
| Q12 | 0.52 | **0.0029** | **0.0053** |
| Q13 | 0.19 | 0.25 | 0.62 |
| Q14 | 0.61 | 0.60 | 0.34 |
| Q15 | 0.07 | 0.46 | 0.47 |
| Q16 | 1.0 | 1.0 | 0.21 |
| Q17 | 0.19 | 1.0 | 0.41 |
| Q18 | 0.26 | 0.27 | 0.98 |

For the results in Table II, "Q1-first" was taken as an example. The value of "Q1-first" was the *p*-value for the answers of Q1, which was calculated with data of the Group (1) and (2), in the first questionnaire *B*. It was observed that no significant difference was observed in all the 18 questions for the first comparison between the Group (1) and (2). The results indicated that no significant difference existed in the knowledge and understandings of the system between the two groups after the explanation of the system. For the third comparisons in Table II, all the participants of the Groups (1) and (2) had experienced both the scenarios (i) and (ii). It was observed that a significant difference existed for Q3, which was related with the ability of the system to recognize objects. During the experiments, the system was designed to be able to recognize vehicles only (car, bus, and truck), and it could not recognize objects including pylons. The participants of the Group (1) with the HMI strongly disagreed that the system could recognize objects other than vehicles, which indicated that they had obtained correct understandings of the system through observing the HMI. For the Q8 and Q9, which were related with the apparent risks (A) and (B), no significant difference was observed. It was considered that the participants of the Group (2) might learn some knowledge of the system by experiencing the apparent risks occurred in the scenarios, even when the HMI was not applied. For the Q2, Q6, Q11 and Q12, they were questions focused on the risks which did not actually appear in the scenarios. It was found that significant differences existed in the third comparison for these questions. The results indicated that the proposed HMI might be effective to continuously present potential risks to drivers and remind them of the potential risks. Finally, the comparatively small number of the participants might be the limitation of the experiment.

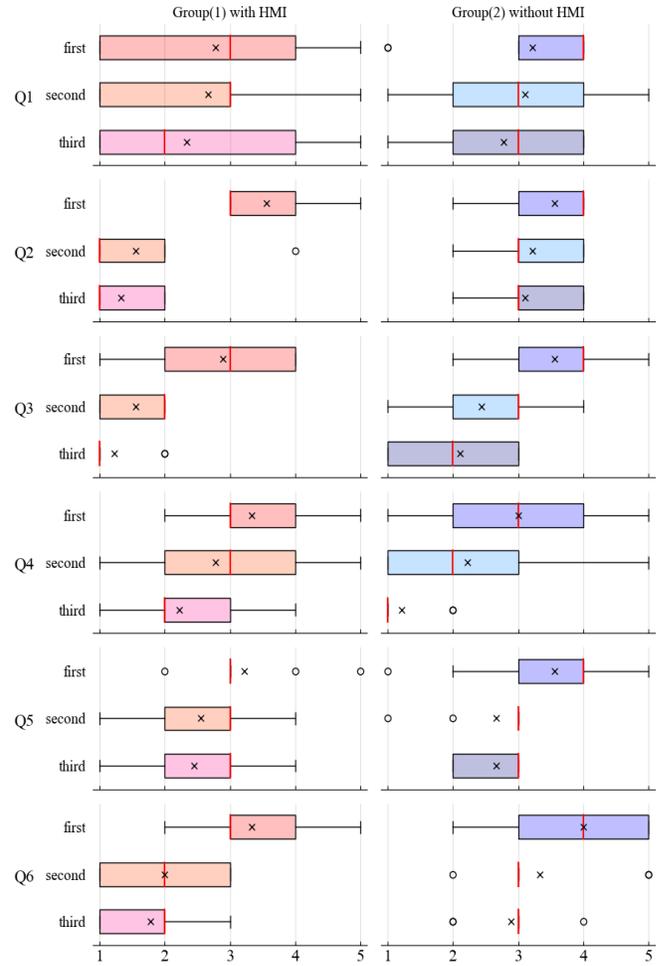

Fig. 12. Results of Q1-Q6 (subjective evaluations on the recognition of the system). The words "first", "second" and "third" mean the order of the questionnaire *B* as shown in Fig. 10.

## V. CONCLUSION

A human machine interface was proposed to improve drivers' understandings of level 2 automated driving systems by presenting the real time image recognition results of the systems to drivers. Driving simulator experiments with 18 participants were performed to analyze the influences while applying the interface. The results verified that the proposed

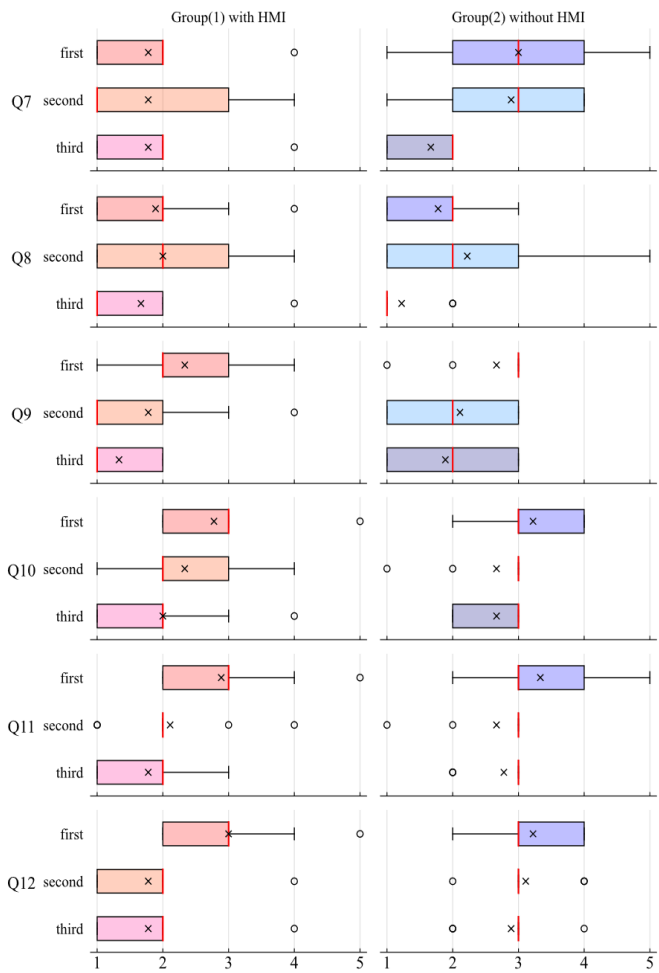

Fig. 13. Results of Q7-Q12 (subjective evaluations on the reaction of the system).

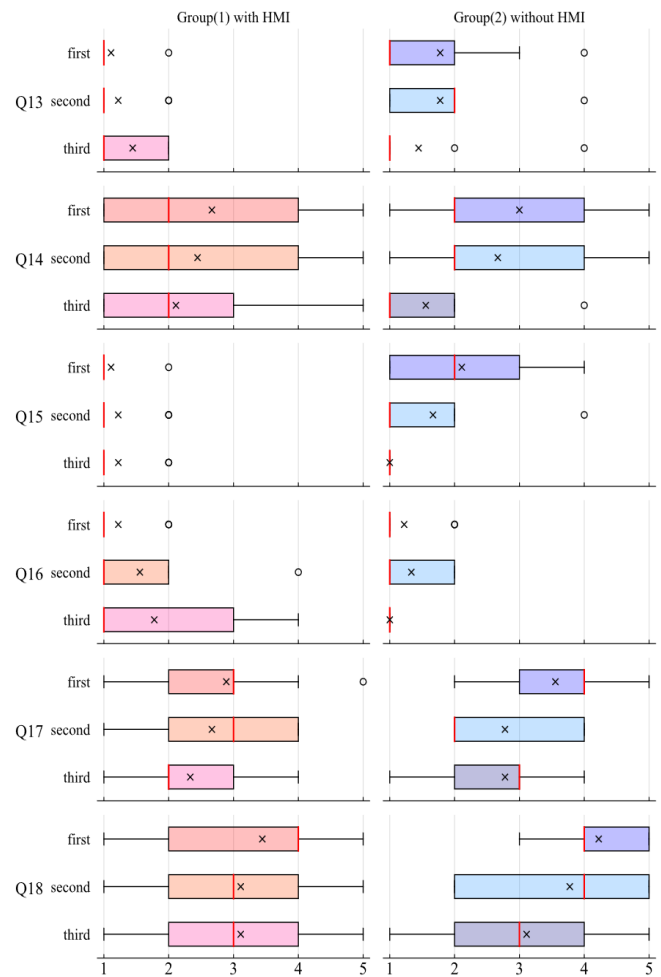

Fig. 14. Results of Q13-Q18 (subjective evaluations on the level of trust for the system).

HMI could improve drivers' understandings of the system and continuously remind them of the potential risks.

ACKNOWLEDGMENT

This research was a part of a national project entitled "HMI and user education for Highly Automated Driving" supported by Cabinet Office, Government of Japan, SIP-adus, and funded by NEDO. The project was also conducted based on Germany-Japan Research Cooperation on Human Factors in Connected and Automated Driving entitled "CADJapanGermany: HF". This work was presented at the Trust calibration for human-automated vehicle interactions workshop (WS02), IV2021.